\documentclass[graybox]{svmult}

\usepackage{mathptmx}       
\usepackage{helvet}         
\usepackage{courier}        
\usepackage{type1cm}        
\usepackage{makeidx}         
\usepackage{graphicx}        
\usepackage{multicol}        
\usepackage[bottom]{footmisc}
\usepackage{hyperref}
\usepackage{amssymb}

\makeindex             

\begin{document}

\title*{Entropy and Transfer Entropy: The Dow Jones and the build up to the 1997 Asian Crisis}
\titlerunning{Entropy and Transfer Entropy: The DJIA and the 1997 Asian Crisis}
\author{Michael S. Harr\'e}
\institute{Centre for Research in Complex Systems \at Faculty of Engineering and IT, Sydney University, Sydney, Australia \\\email{michael.harre@sydney.edu.au}}
%
%
\maketitle

\abstract*{Entropy measures in their various incarnations play an important role in the study of stochastic time series providing important insights into both the {\it correlative} and the {\it causative} structure of the stochastic relationships between the individual components of a system. Recent applications of entropic techniques and their linear progenitors such as Pearson correlations and Granger causality have have included both {\it normal} as well as {\it critical} periods in a system's dynamical evolution. Here I measure the entropy, Pearson correlation and transfer entropy of the intra-day price changes of the Dow Jones Industrial Average in the period immediately leading up to and including the Asian financial crisis and subsequent mini-crash of the DJIA on the 27$^{\mathrm{th}}$ October 1997. I use a novel variation of transfer entropy that dynamically adjusts to the arrival rate of individual prices and does not require the binning of data to show that quite different relationships emerge from those given by the conventional Pearson correlations between equities. These preliminary results illustrate how this modified form of the TE compares to results using Pearson correlation.}

\abstract{Entropy measures in their various incarnations play an important role in the study of stochastic time series providing important insights into both the {\it correlative} and the {\it causative} structure of the stochastic relationships between the individual components of a system. Recent applications of entropic techniques and their linear progenitors such as Pearson correlations and Granger causality have have included both {\it normal} as well as {\it critical} periods in a system's dynamical evolution. Here I measure the entropy, Pearson correlation and transfer entropy of the intra-day price changes of the Dow Jones Industrial Average (DJIA) in the period immediately leading up to and including the Asian financial crisis and subsequent mini-crash of the DJIA on the 27$^{\mathrm{th}}$ October 1997. I use a novel variation of transfer entropy that dynamically adjusts to the arrival rate of individual prices and does not require the binning of data to show that quite different relationships emerge from those given by the conventional Pearson correlations between equities. These preliminary results illustrate how this modified form of the TE compares to results using Pearson correlation.}

\section{Introduction}
\label{sec:1}

One of the most pressing needs in modern financial theory is for more accurate information on the structure and drivers of market dynamics. Previous work on correlations~\cite{mantegna1999hierarchical} has lead to a better understanding of the topological structure of market correlations and mutual information~\cite{harre2009phase} has been used to extend an earlier notion~\cite{vandewalle1998crash,plerou2003econophysics} of a market crash as analogous to the phase-transitions studied in physics. These studies are restricted to static market properties in so far as there is no attempt to consider any form of causation. However, one of the goals of econophysics is to gain a better understanding of market dynamics and the drivers of these dynamics need to be extended to trying to measure causation. This is extremely difficult, strongly non-linear systems such as financial markets have feedback loops where the most recent change in price of equity $a$ influences the price of $b$ which in turn influences the price of $a$. This can make extracting causation relationships exceptionally difficult: the empirical distributions need to accurately reflect the temporal order in which price changes in the equities occur, and the time between these changes is itself a stochastic process. The goal of this paper is to introduce a (non-rigourous) heuristic that addresses these concerns using a modification to the conventional definition of the Transfer Entropy (TE) applied to the intraday tick data of the equities that make up the Dow Jones Industrial Average (DJIA) in the tumultuous build up of the Asian Financial Crisis (AFC) that culminated in the crash of the DJIA on the $27^{\mathrm{th}}$ October 1997. This article is arranged in the following way: Section~\ref{sec:2} introduces the linear Pearson correlations I use as a comparison to the TE introduced in Section~\ref{sec:3} in order to make comparisons and then discuss the results in Section~\ref{sec:5}.

\section{Correlations}
\label{sec:2}

A statistical process generates a temporal sequence of data: $\mathbf{X}_t = \{\ldots, x_{t-1}, x_{t} \}$, $X_t$ is a random variable taking possible states $S_X$ at time $t$, $x_t \in S_X$ and $\mathbf{X}_t^k = \{x_{t-k}, \ldots, x_{t-1}\} \in \{S_X\}^{k-1}$ is a random variable called the {\it k}-lagged history of $X_t$. The marginal probability is $p(X_t)$, the conditional probability of $X_t$ given its $k$-lagged history is $p(X_t|\mathbf{X}^k_t)$ and further conditioned upon the second process $\mathbf{Y}^k_t$ is $p(X_t| \mathbf{X}^k_t, \mathbf{Y}^k_t)$. The Pearson correlation coefficient $r$ between such time series is: 
\begin{eqnarray}
r_t^k & = & \frac{\mathrm{cov}(\mathbf{X}^k_t , \mathbf{Y}^k_t)}{\sigma_X \sigma_Y} \label{stoch_time_series}
\end{eqnarray}
where $\mathrm{cov}(\cdot,\cdot)$ is the covariance, $\sigma_X$ and $\sigma_Y$ are standard deviations and $r_t^k$ is calculated over a finite historical window of length $k$ where in order to calculate the dynamics of $r_t^k$ this window is allowed to slide over the data, updating $r_t^k$ as $t$ progresses. A key issue with data that arrives at irregular or stochastic time intervals and $r^k_t$ is desired is what counts as a co-occurrence at time $t$ of new data. The most common method is to bin the data into equally separated time intervals of length $\delta_t$ and if two observations $x_t$ and $y_t$ occur in the interval $[t-\delta_t, t]$ then $x_t$ and $y_t$ are said to co-occur at time $t$, this approach is used for the correlations calculated in this article. Throughout the change in the log price is the stochastic event of interest: if at time $t$ the price is $p_{t}$ and at time $t'$ it changes to $p_{t'}$ then the stochastic observable is $x_{t'} = \log(p_{t'})-\log(p_{t})$~\cite{stanley2000introduction}, the increment $t' - t$ may be fixed in which case it is labelled $\delta t$ or may dynamically vary, more on this below.

\section{Transfer Entropy}
\label{sec:3}
Transfer Entropy was developed by Schreiber~\cite{schreiber2000measuring} as a rigorous way of measuring the directed transfer of information from one stochastic process to another after accounting for the history of the primary process (see below) for arbitrary distributions. This is a natural extension of Granger Causality, based on covariances rather than information measures, first introduced by Granger~\cite{granger1969investigating} in econometrics and in the case of Gaussian processes Granger causality and Transfer Entropy are equivalent~\cite{barnett2009granger}. Specifically, the entropic measures we are interested in are:
\begin{eqnarray}
\mathbf{H}(X_t) & = & -\mathbf{E}_{p(X_t)}[\log p(X_t)], \\
\mathbf{H}(X_t,Y_t) & = & -\mathbf{E}_{p(X_t, Y_t)}[\log p(X_t, Y_t)], \\
\mathbf{H}(X_t | \mathbf{X}^k_t) & = & -\mathbf{E}_{p(X_t)}[\log p(X_t | \mathbf{X}^k_t)], \\
\mathbf{H}(X_t | \mathbf{X}^k_t,\mathbf{Y}^k_t) & = & -\mathbf{E}_{p(X_t)}[\log p(X_t | \mathbf{X}^k_t,\mathbf{Y}^k_t)],
\end{eqnarray} 
where $\mathbf{E}_{p(\cdot)}[\cdot]$ is the expectation with respect to distribution $p(\cdot)$. The mutual information between two stochastic time series $\mathbf{X}_t$ and $\mathbf{Y}_t$ is: 
\begin{eqnarray}
\mathbf{I}(\mathbf{X}_t;\mathbf{Y}_t) &\equiv & \mathbf{H}(\mathbf{X}_t) -  \mathbf{H}(\mathbf{X}_t | \mathbf{Y}_t) \; = \; \mathbf{H}(\mathbf{Y}_t) -  \mathbf{H}(\mathbf{Y}_t | \mathbf{X}_t)
\end{eqnarray}
with a finite data window of length $k$ this is the information theoretical analogue of $r^k_t$ and the $k$-lagged {\it transfer entropy} (TE) from the {\it source} $\mathbf{Y}$ to the {\it target} $\mathbf{X}$ is:
\begin{eqnarray}
\mathbf{T}^k_{Y\rightarrow X} & \equiv & \mathbf{H}(X_t | \mathbf{X}^k_t) - \mathbf{H}(X_t | \mathbf{X}^k_t,\mathbf{Y}^k_t).
\end{eqnarray} 
$\mathbf{T}^k_{Y\rightarrow X}$ measures the degree to which $X_t$ is disambiguated by the $k$-lagged history of $Y_t$ beyond that to which $X_t$ is already disambiguated by its own $k$-lagged history. This work presents recent developments in TE~\cite{barnett2012transfer}, information theory and the `critical phenomena' of markets~\cite{harre2009phase}, and adds new results for real systems to the recent success in using it as a predictive measure of the phase transition in the 2-D Ising model~\cite{barnett2013information}. The implementation of TE used in this work was done in Matlab using~\cite{lizier2014jidt}.

\subsection{Transfer entropy without binning}
\label{sec:4}

The most common and direct method of calculating any of $r^k_t$, $\mathbf{I}(\mathbf{X}_t;\mathbf{Y}_t)$ or $\mathbf{T}^k_{Y\rightarrow X}$ is to use discrete time series data. This is made possible either by the nature of the study itself where discrete time steps are inherent or through post-processing of the data by binning it into a discrete ordered sequence. However, a lot of interesting data, including intra-day financial markets data, is inherently unstructured and binning the data loses some of the temporal resolution and obfuscates the relationship between past and future events making causal relationships difficult to establish, so an alternative is proposed that addresses these issues.

I define a modified form of $\mathbf{T}^k_{Y\rightarrow X}$ by first redefining the stochastic time series in order to capture the continuous nature of the price arrival process. With $t$ and $t' \in \mathbb{R} > 0 $ where 0 is taken as the start of trading on any given trading day and $\{t_i\}$ and $\{t'_j\}$ are the finite sequence of times at which the (log) price changes for two different equities during that day. Define the arrival indices of time series of length $I$ and $J$ as $\{i \leq I\} \in \mathbb{N}$ and $\{j \leq J \} \in \mathbb{N}$. Now there are two finite sequences of price changes on a single trading day $d$: $\{X^d(t_i)\}$ and $\{Y^d(t'_j)\}$. The entropy of $\{X^d(t_i)\}$ conditioned on its most recent past value is:
\begin{eqnarray}
\mathbf{H}(X^d(t_i) | X^d(t_{i-1})) & = & -\mathbf{E}_{p(X^d)}\big[\log(p(X^d(t_i) | X^d(t_{i-1}) ) \big], \;\; i>1.
\end{eqnarray}
An equivalent definition for the entropy conditioned on the most recent past of both $\{X^d(t_i)\}$ and $\{Y^d(t'_j)\}$ is:
\begin{eqnarray}
\mathbf{H}(X^d(t_i) | X^d(t_{i-1}),Y^d(t'_{j-1})) & = & -\mathbf{E}_{p(X^d)}\big[\log(p(X^d(t_i) | X^d(t_{i-1}),Y^d(t'_{j-1}) ) \big]
\end{eqnarray}
where $i,j >1$ and $t'_{j-1}$ is the minimum value such that, for a given $t_i$: $(t_i - t'_{j-1})>0$. This modified definition of the TE (for the rest of this article this is simply referred to as {\it the} TE) is:
\begin{eqnarray}
\mathbf{\overline{T}}_{Y^d\rightarrow X^d} & \equiv & \mathbf{H}(X^d(t_i) | X^d(t_{i-1})) - \mathbf{H}(X^d(t_i) | X^d(t_{i-1}),Y^d(t'_{j-1})). \label{mod_TE}
\end{eqnarray}

\begin{figure}
\begin{center}\label{concept}
\includegraphics[width=0.9\hsize]{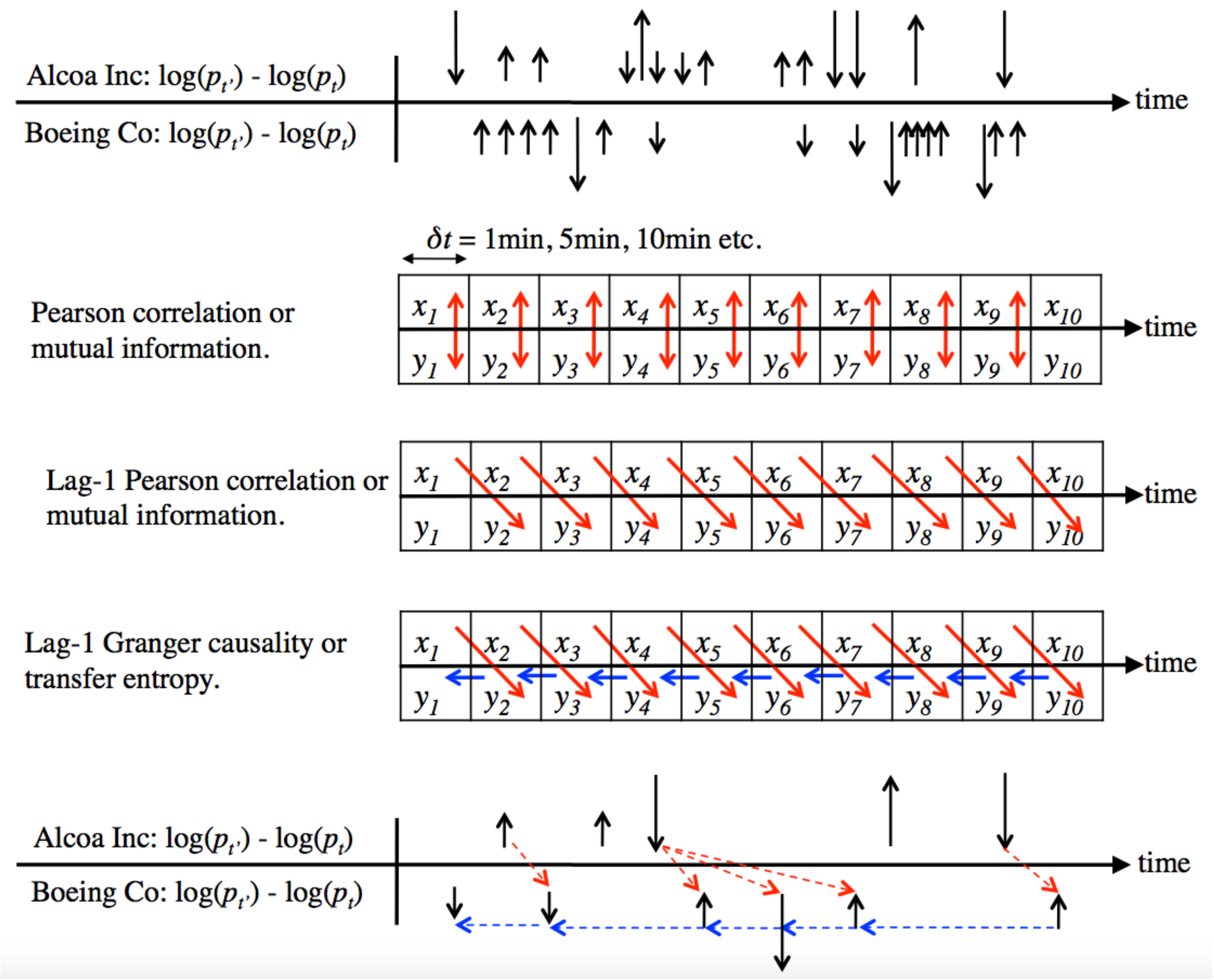}
\caption{A representation of different measures of `instantaneous' and `lagged' relationships between stochastic time series data.}
\end{center}
\end{figure}

The relationship between this and other measures is illustrated in Figure 1. The first row shows the log price changes for two equities (Alcoa and Boeing) as a stochastic time series with an irregular arrival rate. The black arrows indicate the direction and magnitude of the log price changes. The second row shows the changes in prices binned into time intervals of width $\delta t$ so that changes that occur in the same time interval are considered co-occurring. In the third row is the lag-1 Pearson correlations or lag-1 mutual information, the causal direction of correlations is implicit in the time ordering of the bins, hence the arrows point forward in time. This does not account for the shared signal between $x_{t-1}$ and $y_{t-1}$. The fourth row shows the lag-1 Granger causality or transfer entropy, the signal driving $y_t$ is $x_{t-1}$ after excluding the common driving factor of $y$'s past: $y_{t-1}$. Red arrows indicate the measured signal from the source (Alcoa) to the target (Boeing) and blue arrows indicate $y$'s signal that is being removed. Fifth row (fewer price changes shown for clarity): An alternative way to calculate the TE. Choose the target time series (in this case Boeing) and condition out the most recent previous price change in Boeing and then use only the most recent change in Alcoa as the source signal. Note that some Alcoa price signals are missed and some are used more than once and that price changes will rarely co-occur.

The definition of Equation~\ref{mod_TE} has a number of appealing properties: 
\begin{itemize}
\item Using a fixed interval in which the price at the beginning is compared with the price at the end of the interval conflates signals that may occur before or after another signal but arrives during the same binning interval, thereby mixing future and past events in the measured relationships between bins.
\item Similarly, multiple price changes within $\delta t$ may net to zero change and so some price signals are missed.
\item As bin sizes get smaller they are less statistically reliable as fewer events occur within each bin, equally as bin sizes get larger there are fewer bins per day, thereby also reducing the statistical reliability.
\item Over the period of a single day, for each bin size the number of total bins is: $\delta t = 30$ minutes: 13 bins/day,  $\delta t = 1$ minute: 390 bins/day, whereas the raw data may have 50 to 5,000+ price changes in a day.
\end{itemize}

The proposed heuristic for the TE introduced above addresses some of these shortcomings but not without introducing some other issues. First, it will always condition out the most recent price change information in the target equity (Boeing in Figure~1) and so uses every bit of relevant information in the target time series. It also uses the most recent price change from the source time series, however it will sometimes miss some price changes or repeatedly count the same price changes (see bottom of Figure~1). This is good if we are interested in the most recent price signals and in financial markets this is the case. It also reflects the dynamical nature of the time series, as the inter-arrival times may vary from day to day or between equities no new $\delta t$ needs to be defined, it will always use only the most recent information in both the source and the target time series. The most significant shortcoming is that this TE assumes there is no information being carried by the inter-arrival time interval and it is not clear that some of the theoretical foundations on which the original TE is based necessarily hold, from this point of view this method of calculating the TE is currently only a heuristic and the results presented here are for the moment qualitative in nature. 

\section{Empirical Results}
\label{sec:5}

The AFC began in Thailand in July 1997 with the devaluation of the Thai currency (the Bhat) and the crisis rapidly spread throughout South East Asia, ultimately resulting in the October 27 ``mini-crash'' of the DJIA, losing around $7\%$ on the day which was at the time the largest single day points drop on record for the DJIA, for a review of the crisis see~\cite{radelet1998onset} and the top plot of Figure~2. Note that the entropy measurements shown illustrate that some care needs to be taken when comparing simple systems with data from real `complex systems': the increase in the entropy of the DJIA on the $24^{\mathrm{th}}$ of June looks like what might be described as a `first order' phase transition as studie din complex systems~\cite{li1990transition}, but it is almost certainly caused by the rescaling of price increments on the New York Stock Exchange\footnote{For details see: \url{http://www1.nyse.com/nysenotices/nyse/rule-changes/detail?memo_id=97-33}}.

\begin{figure}
\begin{center}
\includegraphics[width=0.9\hsize]{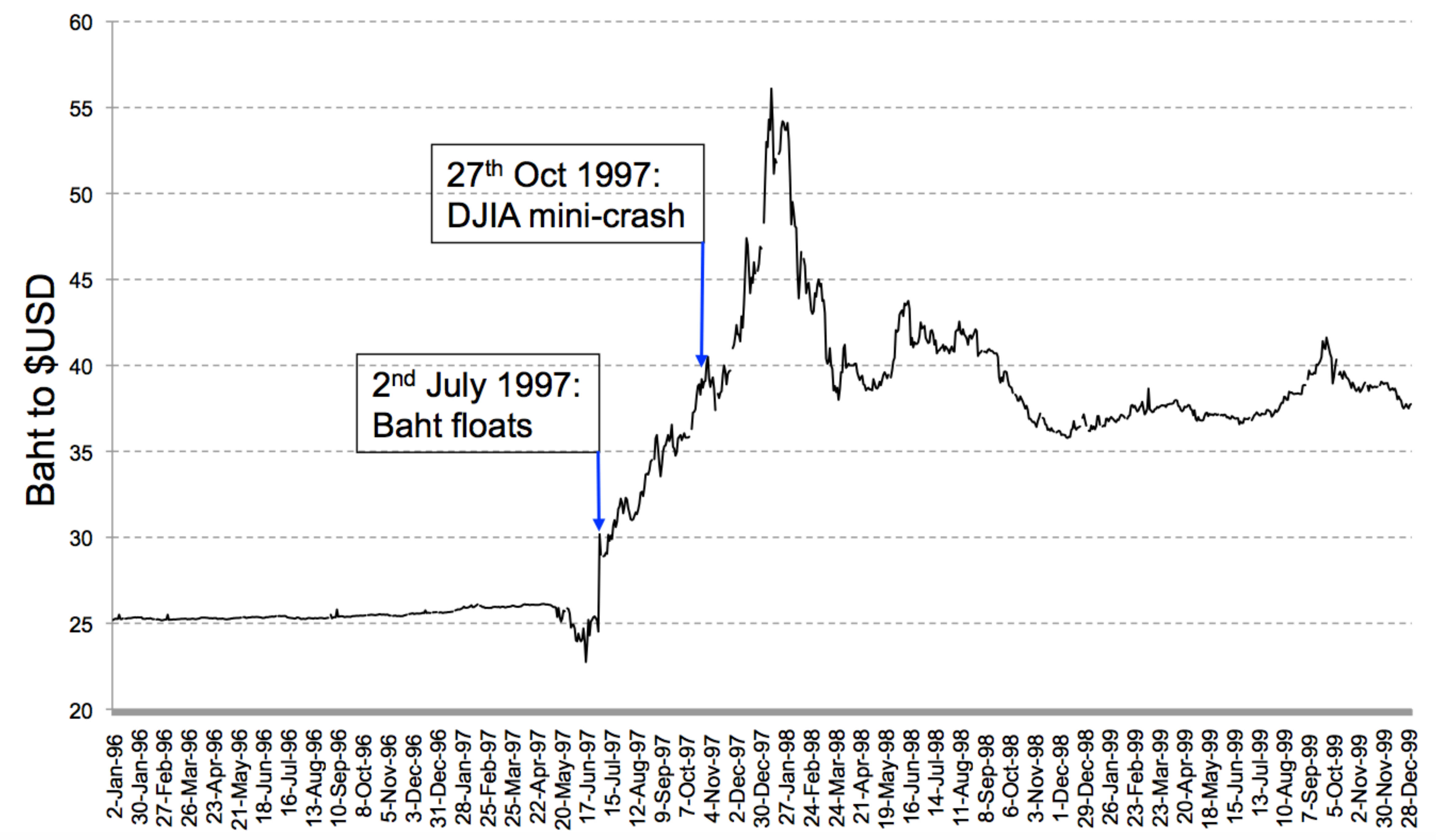}
\includegraphics[width=0.9\hsize]{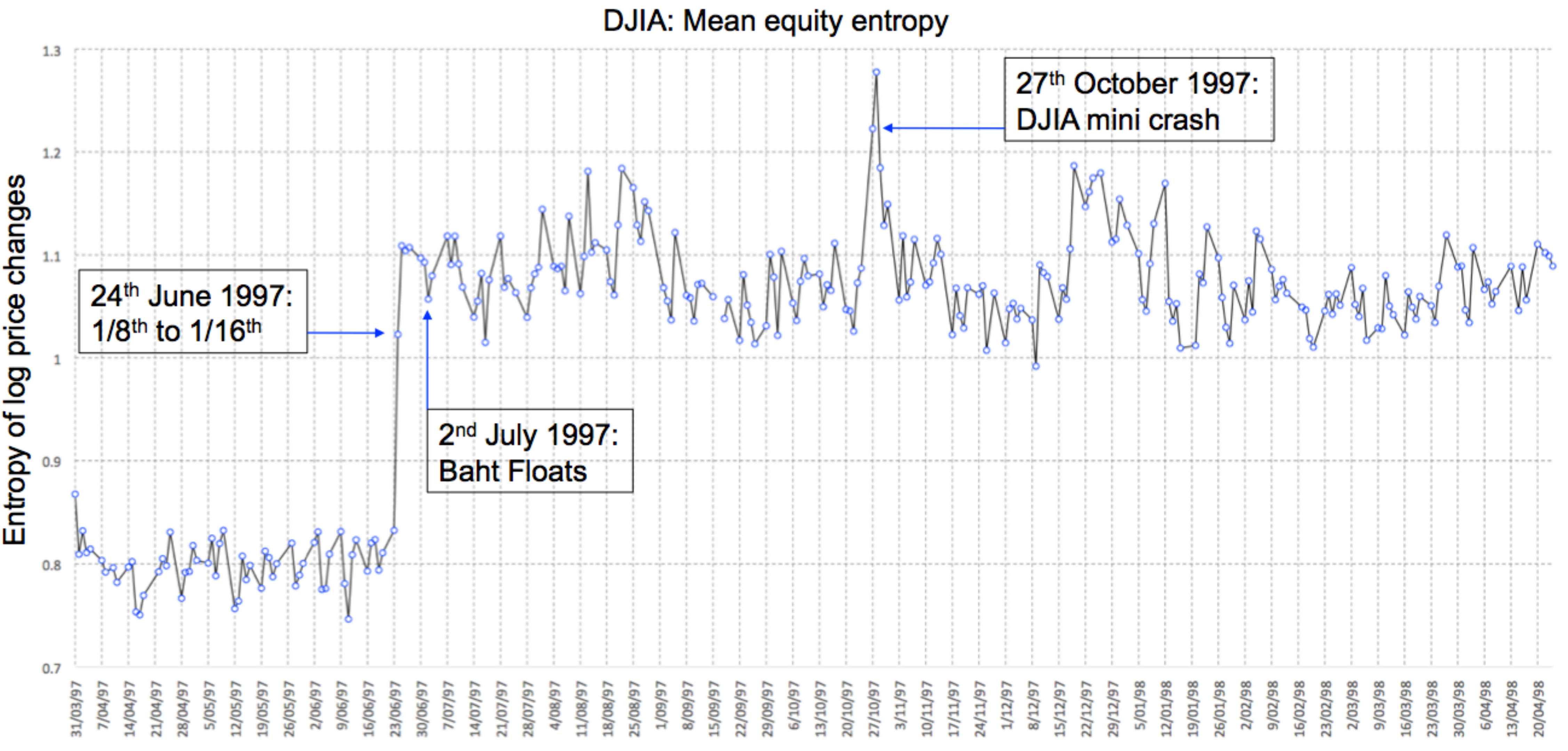}
\caption{The AFC and its key components. Top plot: the AFC is thought to have begun as the Baht was allowed to float against the US dollar on the $2^{\mathrm{nd}}$ of July 1997. The crisis contagion spread through the asian markets ultimately leading to the mini-crash of the DJIA on the $27^{\mathrm{th}}$ October 1997. Bottom plot: On the $24^{\mathrm{th}}$ of June 1997 the New York Stock Exchange changed its minimum incremental buy/sell price from $1/8^{\mathrm{th}}$ of a dollar to $1/16^{\mathrm{th}}$ of a dollar, causing the entropy of the price changes to shift suddenly and permanently, but not influencing the DJIA index itself. The crash on the $27^{\mathrm{th}}$ October 1997 is seen as the second largest peak in the entropy, the largest being the $28^{\mathrm{th}}$ of October.}
\end{center}
\label{fig:2}
\end{figure}

This rescaling did have an interesting impact on the TE though, as can be seen in Figure~3. Prior to the $24^{\mathrm{th}}$ of June there is considerable structure in the TE measure (warm colours denote high TE values, cooler colours denote lower TE values), however all signals drop off significantly immediately after this date although much of the structured signal eventually returns (not shown). The most notable signals are equities that act as targets of TE for multiple other equities, seen as yellow vertical strips indicating that many equities act as relatively strong sources of TE for a single equity: AT\&T (equity 26), Wall Mart (equity 30) and McDonalds (equity 31) stand out in this respect. Notable single sources of TE are less obvious but Cocoa Cola and AT\&T (equities 19 and 26) show some coherent signals indicated by multiple red points loosely forming a horizontal line. It is intriguing to note that the Pearson correlations showed no similar shift on the $24^{\mathrm{th}}$ of June (not shown) while conversely in Figure~4 the mini-crash on the $27^{\mathrm{th}}$ October 1997 (day 64) there is a clear signal that the DJIA equities are significantly more correlated with no corresponding increase in the TE on that day (not shown) despite the general turmoil of the markets, as seen by significant fluctuations in the correlations on nearby days. 

\begin{figure}
\begin{center}
\includegraphics[width=0.9\hsize]{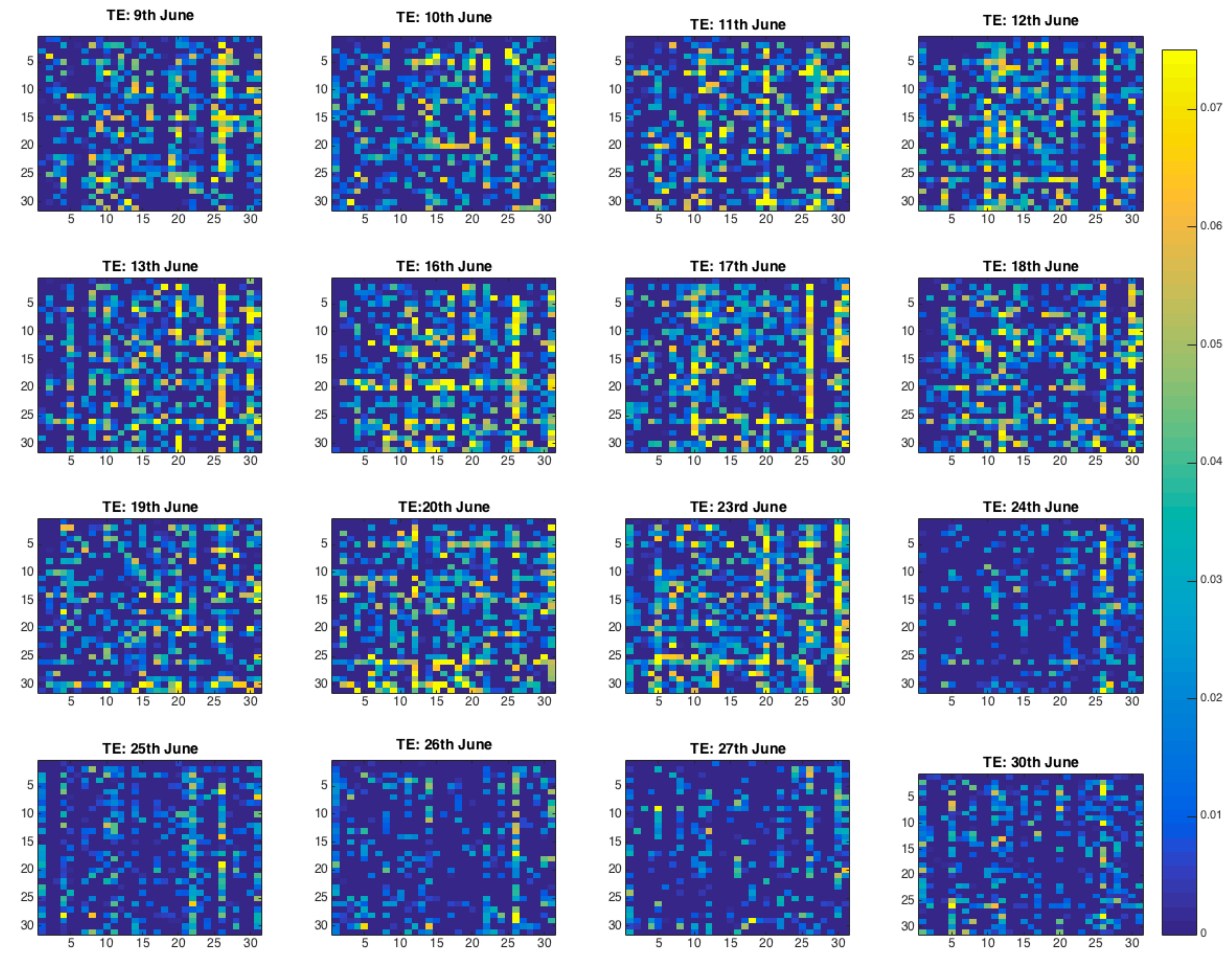}
\includegraphics[width=0.9\hsize]{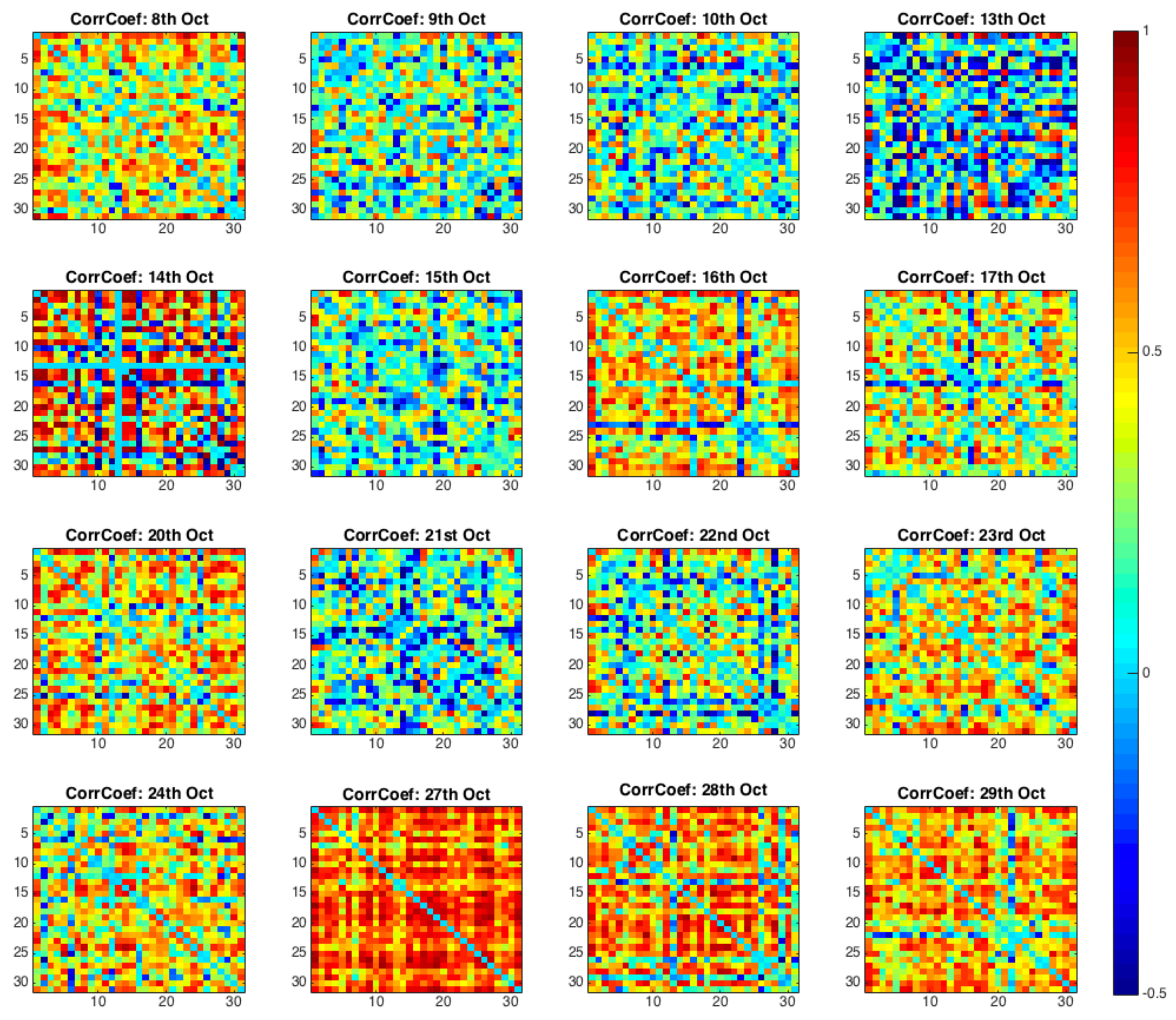}
\caption{{\bf Top}: The TE from one DJIA equity to another equity indexed from 1 to 31. Index 1 = the DJIA, vertical axis is the {\it source} equity, horizontal axis is the {\it target} equity. The $24^{\mathrm{th}}$ of June 1997 clearly stands out as the first day of a substantive reduction in the TE between equities. {\bf Bottom}: The Pearson correlation for the DJIA data binned using $\delta t = 30$ minutes. The market crash on the $27^{\mathrm{th}}$ October stands out during a turbulent time in the market's dynamics.}
\end{center}
\label{TE_DJIA}
\end{figure}

In Figure~4 it is shown that the average TE across all equities is quite stable except for the drop occurring at the time of the change in minimum price increments on the $24^{\mathrm{th}}$ June. A simple shuffling test~\cite{kwon2008information} estimates the TE for unrelated data to be approximately 0.02 nits on average (see the dashed lines, randomly sampled before and after the drop on day 61) but note that numerical estimations of TE are difficult so the TE sometimes drop below zero. This suggests that {\it on average} the TE across the DJIA is close to negligible but that some equities clearly have TE values {\it significantly exceeding} the 0.02 nits level, as shown by the blue line values. The largest peak in the Maximum TE plot occurs 6 days after the Dow crashes and is from the Disney equity to the McDonalds equity.

\begin{figure}
\begin{center}
\includegraphics[width=1\hsize]{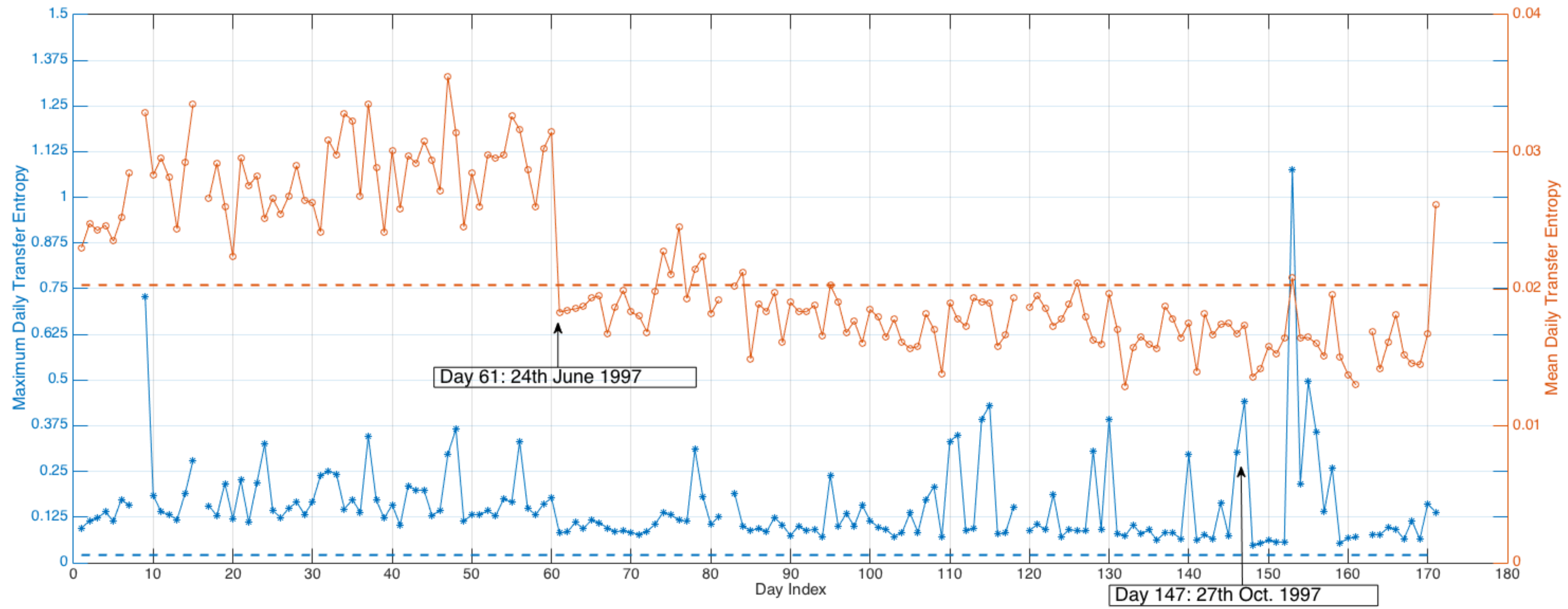}
\caption{Maximum (blue, lower line measured from the left axis) versus average (orange, upper line measured from the right axis) TE. Shuffled average of TE $\simeq 0.02$ are blue dashed for left axis and orange dashed for right axis.}
\end{center}
\label{mean_vs_av}
\end{figure}

\begin{figure}
\begin{center}
\includegraphics[width=0.9\hsize]{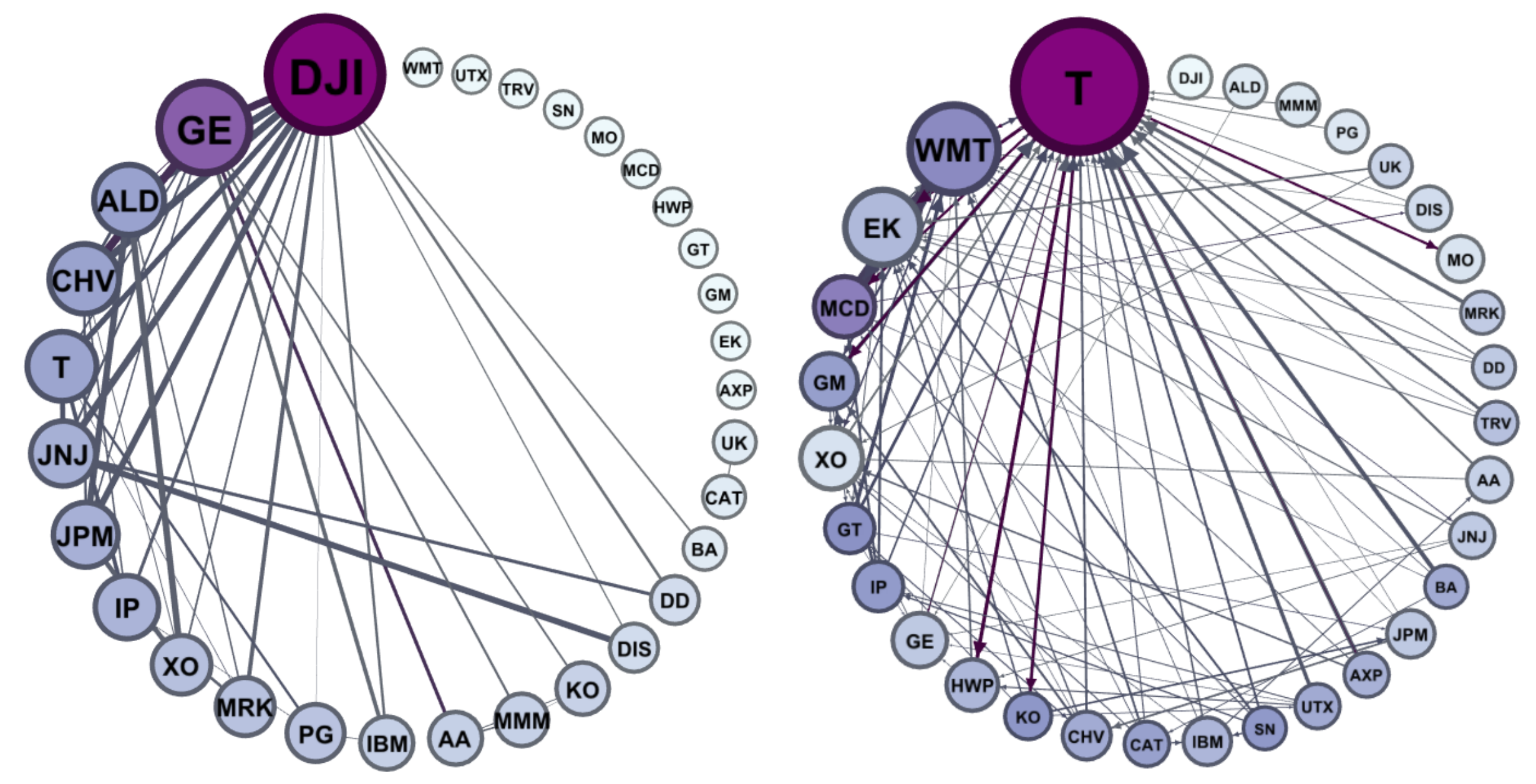}
\caption{The Pearson correlation (left, undirected links) compared to TE (right, directed links) network of relationships for a typical trading day ($16^{\mathrm{th}}$ June 1997).}
\end{center}
\label{gephi}
\end{figure}

Finally in Figure 5 is plotted two networks of relationships between the equities based on Pearson correlations and TE. The Pearson correlation network is ordered counterclockwise according to the total link weight of each equity and a link was included if its correlation was greater than 0.4. The TE network is ordered counterclockwise by total link weight, the colour represents the total weight of incoming links and the node size represents the total weight of outgoing links and a link was included if its TE was greater than 0.05 nits. Thresholds were chosen such that 10\% of all links in each network are included. The most notable differences between these networks is the changes in the relative importance of the individual equities. The overall DJIA index (DJI) is significantly correlated with other equities whereas this index is the {\it least} significant node in the TE network. Similarly, Walmart (WMT) is very well connected in the TE network but it is the least relevant node in the Pearson correlation network.

These are preliminary results using the comparatively small dataset of the 30 equities that make up the DJIA and will need to be confirmed on other indices and other crashes. There is one very significant point that comes out of this study: The driver of correlations between equities in financial markets is not necessarily the changes in the prices of other equities. This is true in the sense that changes in transfer entropy may leave correlations unchanged and changes in correlations are not necessarily driven by changes in transfer entropy. The former is a consequence of the top plot of Figure~3 (the plots showing the lack of change in correlations is not shown due to space limitations), the latter is a consequence of the lower plot Figure~3 for the Asian crisis crash (the plots showing the lack of change in transfer entropy are not shown). However, in the case of the Asian crash, the transfer entropy significantly peaked several days after the crisis but the significance of this is not clear from the data. This result is not peculiar to trading days in which known `significant' events have occurred. Figure~5 shows an ordinary trading day in which the DJIA index plays a significant role in the correlation structure (left plot) but this relationship vanishes for the transfer entropy structure (right plot), compare for example the position of Walmart (WMT) in the two plots. In fact there appears to be very little relationship between strongly correlated equities and those that `transfer' high values of entropy.

One of the goals of this work was to explore the analogy between phase transitions in statistical physics and market crashes in finance. Although recent work on precursors to phase transitions in physics has shown that it is a peak in a global measure of TE acts as precursor~\cite{barnett2013information}, it is interesting that peaks in Pearson correlations are not necessarily coincidental with peaks in TE for financial markets suggesting that it is not the transfer of entropy between equities within the DJIA that is driving the correlations but some signal external to the market. The results in~\cite{barnett2013information} suggest that if the DJIA mini-crash was analogous to the second order phase-transition in the Ising model then peaks in the pairwise TE, mutual information and Pearson correlation~\cite{matsuda1996mutual} would be observed at the crash. However, in this and earlier studies only peaks in Pearson correlations and mutual information have so far been established during a market crash requiring verification and opening up a number of interesting questions for further work.


\end{document}